\documentclass[aps,twocolumn,amsmath,amssymb,floatfixng,superscriptaddress,longbibliography,numerical]{revtex4-1}
\pdfoutput=1
\usepackage[dvips]{graphics}
\usepackage{bm}
\usepackage{epsfig}
\usepackage{enumerate}
\usepackage{subfigure}
\usepackage{color}
\usepackage{graphicx}
\usepackage{hyperref}
\hypersetup{
    pdfnewwindow=true,      % links in new window
    colorlinks=true,       % false: boxed links; true: colored links
    linkcolor=magenta,          % color of internal links
    citecolor=blue,        % color of links to bibliography
    filecolor=blue,      % color of file links
    urlcolor=blue        % color of external links
}

\newcommand{\bea}{\begin{eqnarray}}
\newcommand{\eea}{\end{eqnarray}}
\newcommand{\beq}{\begin{equation}}  
\newcommand{\eeq}{\end{equation}}
\newcommand{\eps}{\epsilon}

\begin{document} 
\title{Propagation of light through amplifying honeycomb photonic lattice}

\author{SK Firoz Islam}
\affiliation{Department of Applied Physics, Aalto University, P.~O.~Box 15100, FI-00076 AALTO, Finland}

\author{Pascal Simon}
\affiliation{Universit{\'e} Paris-Saclay, CNRS, Laboratoire de Physique des Solides, 91405, Orsay, France}

\author{Alexander A. Zyuzin}
\affiliation{Department of Applied Physics, Aalto University, P.~O.~Box 15100, FI-00076 AALTO, Finland}
\affiliation{Ioffe Physical--Technical Institute,~194021 St.~Petersburg, Russia}

\begin{abstract}
We consider light propagation through a ballistic amplifying photonic honeycomb lattice below the lasing threshold.
Two sublattices of the system are formed by the wave-guides with different complex dielectric permittivities, which results in the non-Hermitian Dirac equation for electromagnetic field. 
We reveal that there exists a critical length of the amplifying region for which the photonic lattice exhibits an amplifier to generator transition. The transmission and reflection probabilities at the normal angle of incidence are strongly enhanced at a critical length of the system. We also comment on the sensitivity of amplification to the direction of incident light and the thickness of the amplifying region.
\end{abstract}
\maketitle

%-------------------------
\section{Introduction}
%-------------------------
A two-dimensional (2D) photonic lattice (PhL) \cite{PhysRevB.44.8565, PhysRevLett.100.013904} has emerged as a versatile platform in engineering the optical analogs of most of the interesting quantum phenomena related to the non-trivial band topology occurring in 2D condensed matter systems \cite{Haldane, Volovik,Review_photonics_2018,
Alexander}.
In particular, it was first noted that a periodic array of parallel waveguides, forming an intersection of triangular lattice, can split the dispersion curves by an absolute gap in frequency \cite{PhysRevB.44.8565}. Tamm surface waves were shown to exist at frequencies within the band gap for certain lattice terminations \cite{Robertson}. It was later shown that photonic crystals might have Dirac points in the band structure at certain frequency \cite{PhysRevB.44.8565, PhysRevLett.100.013904}. An analog of the quantum Hall effect and of the anomalous quantum effect have been proposed by utilizing the interplay of broken spatial inversion and time reversal symmetries and the band-structure topology at the Dirac point \cite{Onoda, PhysRevLett.100.013904}. 
Reviews on the optical analogs of the electronic band structure and transport properties can be found in \cite{Beenakker_analogs, lu2014topological, Konotop_RevModPhys,torres,Review_photonics_2018}.

Let us specify that the scattering phenomena in a Dirac material such as graphene differs fundamentally from the electronic system described by a quadratic band structure \cite{katsnelson2006chiral, Klein_review}. In certain regime of parameters, the Dirac electron can fully tunnel through the potential barrier without any reflection, a phenomenon known as Klein tunneling. A photonic analog of  Klein tunneling
 was also investigated in PhL and a transition from unit transmission to fully reflection of electromagnetic (EM) wave with respect to the band structure deformation was found \cite{PhysRevLett.104.063901}. The effects of the interfacial coupling between air media and a PhL on the EM wave transmission through a PhL was also investigated
\cite{PhysRevA.75.063813}. A possibility of negative refractive index has recently been predicted by studying  EM wave transmission though driven-dissipative background \cite{PhysRevA.96.013813}.

However, no detailed studies of the Klein tunneling process of EM wave through a 2D honeycomb PhL with amplifying background have been reported so far.
The effect of such background on the EM wave transmission through photonic crystals has already attracted intense research interests after the theoretical proposals of parity-time ($\mathcal{PT}$) symmetric laser absorber \cite{PhysRevA.82.031801} and laser as coherent perfect absorber \cite{PhysRevLett.105.053901}.
This has further been boosted by a series of works which are reviewed for example in \cite{Konotop_RevModPhys}.
Very recently, several artificial techniques of imparting non-Hermiticity in the $2$D honeycomb PhL have been reported which opened up a possibility of experimental realization of Klein tunneling through loss or gain medium  \cite{Konotop_RevModPhys, Review_photonics_2018,torres}.

Here we take this advantage to investigate the EM wave propagation through an amplifying PhL at the vicinity of Dirac point in the spectrum of wave. The amplification background is attributed to the imaginary part of the dielectric constants of the waveguides. We find that there exists a critical length of the amplifying region for which the transmission and reflection probabilities diverge, {\it  i.e.}, resonance does occur. Above the threshold, there exists a generator solution. 
For the limiting case of the scattering problem for an imaginary delta-function potential, the resonance condition is related to a critical strength of the barrier. 
We also comment on the effect of an imaginary gauge field on the transmission probability, which is known to exponentially suppress 
transmission in preferred directions of light propagation \cite{Longhi_robust, PhysRevB.92.094204}.

%----------------------------------------------------------------------
\section{Non-Hermitian Hamiltonian.}
%----------------------------------------------------------------------
We consider a 2D PhL in $x-y$ plane formed out of cylindrical waveguides aligned parallel to the $z$-axis. These waveguides are arranged in a way to form a hexagonal cross section, mimicking the graphene geometry by replacing each sublattice point by a waveguide. The 2D cross section of the PhL is described by a frequency dependent dielectric constant $\varepsilon_{\omega}(x,y)$ which is periodic in the $x-y$ plane.
We consider a simple set up which consists of two regions with positive and real dielectric permittivity (regions I and III) separated by an amplifying region II of length $L_x$. The interior of waveguides in the region-II has complex dielectric permittivity, which describes the response of oscillating dipoles of frequency $\omega_0$.

%%%%%%%%%%%%%%%%%%%%%%%%%%%%%%%%%%%%%%%%%%%%%%%%%%%
\begin{figure}[t]
\centering
\includegraphics[height=3.8cm,width=\linewidth]{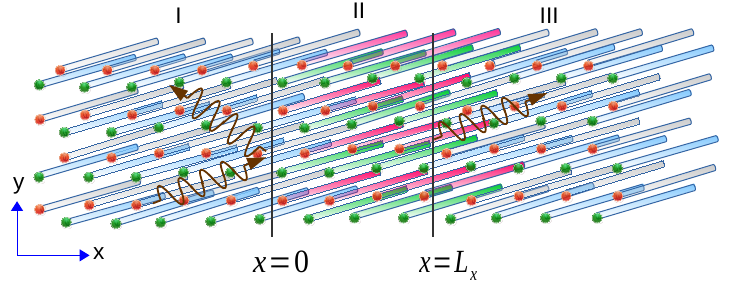}
 \caption{Schematic sketch of the device. Two different colors are used to denote $A$ and $B$ sublattice (here waveguides).
The wave guides in the region-II, of length $L_x$, are embedded into a non-Hermitian background for which different colored are used. The regions-I and III are considered to be identical and quite longer than region-II. In the region-II, the waveguides are filled with media described by a complex dielectric permittivity. 
}
\label{device}
\end{figure}
%%%%%%%%%%%%%%%%%%%%%%%%%%%%%%%%%%%%%%%%%%%%%%%%%%%

Here, in order to derive the wave-equation for the propagation of the EM field in a photonic honeycomb lattice, we
follow Ref. [\onlinecite{PhysRevB.44.8565}] and  Ref. \cite{PhysRevB.98.165129} for explicit derivations.
Let us start with the Helmholtz equation for the $z$-component of the EM field $ \sim E_z(x,y) e^{-i\omega t}$, passing through the PhL, as
\begin{equation}
 \left[\varepsilon_{\omega}^{-1}(x,y)\boldsymbol{\nabla}^2 + \omega^2/c^2\right]E_{z}(x,y)=0,
\end{equation}
where $c$ is the speed of light. In the regions I and III, the periodic arrangement of waveguides allows to employ the Bloch 
theorem. The field component $E_z$ on the honeycomb lattice can be written using irreducible singlet and doublet representations for two sets of inequivalent
corners of the hexagonal first Brillouin zone. The doublet states form a Dirac point, while non-degenerate singlet states are well separated in frequency from the Dirac
points and will be ignored. One can show that the field satisfies the Dirac like eigenvalue equation (see \cite{Review_photonics_2018} for a review)
\begin{equation}\label{wave_2}
\left(\Omega +i v\boldsymbol{\sigma}\cdot \boldsymbol{\partial}\right)E=0,
\end{equation}
where $\Omega = (\omega^2-\omega_D^2)/\omega_D$ in which $\omega_D$ is the frequency of the band touching point, $v$ is the velocity term, $\sigma_{x,y,z}$ are the Pauli matrices. 
Note that the above Dirac-like band structure appears for both modes \cite{Alexander}: the  TE mode ($E_z=0,H\ne 0$) and the TM mode ($E_z\ne 0,H_z=0$), but at different frequency ($\omega_D$) with different velocity parameter ($v$)

The optical medium with loss and gain background might be described by a non-Hermitian Dirac Hamiltonian, whose unique feature is the emergence of exceptional points or lines, where the complex frequency eigenvalues coalesce, \cite{PhysRevLett.80.5243, Berry, Konotop_RevModPhys, PhysRevX.9.041015}. In the region-II, the complex dielectric constant inside the waveguides results in complex additive terms
in the wave equation as (for details of the derivation see for example \cite{PhysRevB.98.165129})
\begin{equation}\label{non_Hermitian_Dirac}
\left(\tilde{\Omega} + iv\boldsymbol{\sigma}\cdot \boldsymbol{\partial}
+i\Gamma + i\gamma\sigma_z +i\epsilon \sigma_x \right) E=0,
\end{equation}
where $\tilde{\Omega} = (\omega^2- \tilde{\omega}_D^2)/\tilde{\omega}_D$ denotes the frequency of the band touching point in region-II, $\tilde{\omega}_D\neq \omega_D$. The negative (positive) values of $\Gamma$
determine the overall gain (loss) in the system, while $\gamma$ describes the difference in the amplification (dissipation) between the two sublattices. Note that while the imaginary term $i\gamma\sigma_z$ violates both spatial inversion symmetry as well as time reversal symmetry, the $i \Gamma$ breaks only the latter. The last term, $i\eps\sigma_x$, is the imaginary gauge field which might stems from an imaginary magnetic flux, \cite{Longhi_robust}.
In what follows we  consider an amplifying system $\Gamma < 0$ with $|\Gamma| > |\gamma|, |\epsilon|$ and shall comment further on the other cases.

\section{Scattering through amplifying media} Let us now discuss how the amplifying background affects the ballistic transmission of a EM
wave through a PhL. We consider the standard scattering problem by seeking for a transmission and reflection probabilities of the incident radiation in the geometry, as shown in Fig.~\ref{device}. We shall assume that the wave-vector parallel to the interfaces is conserved as well the radiation frequency at the Dirac point, namely $\omega = \tilde{\omega}_{D}$. 
The problem is analogous to the scattering problem through a rectangular potential barrier in graphene \cite{katsnelson2006chiral, Klein_review}, except that the real potential barrier is replaced by a complex one.

The eigenstates in regions I and III corresponding to Eq. \eqref{wave_2} can be written, respectively, as
\begin{align}\nonumber
&E_1(x,y) = [n\chi_{+}e^{i k_x x} + r \chi_{-}e^{-ik_x x}]e^{ik_y y},\\
&E_3(x,y) = t\chi_{+}e^{i (k_x x+k_y y)},
\end{align}
where the spinor part is given by $\chi_{\pm}=[1, v(\pm k_x+ik_y)/\Omega]^{T}$. The reflection and transmission amplitudes are denoted by $r$ and $t$.
The wave-vector can be parametrized with the angle of incidence as $vk_x= |\Omega|\cos\theta$
and $vk_y= |\Omega|\sin\theta$. 

Inside the amplifying region II, the propagation of light is dominated by the evanescence mode at $\omega = \tilde{\omega}_{D}$. 
The eigenstate corresponding to Eq. \eqref{non_Hermitian_Dirac} can be written as
\begin{equation}
E_2(x,y)= [a \psi_{+}e^{-\kappa x} + b \psi_{-}e^{\kappa x}] e^{ik_y y}e^{-\epsilon x/v}, 
\end{equation}
where the spinor part of
the solution is given by $\psi_{\pm}=[1, v(\pm \kappa+k_y)/(\Gamma-\gamma)]^{T}$ and $\kappa$ can be determined from $v\kappa = \sqrt{v^2k_y^2+\Gamma^2-\gamma^2}$. The effect of $\gamma$ on  backscattering at the normal angle of incidence can be seen from the spinor structure of the solution.

Note that unlike the case of a Hermitian scattering problem, here the energy of the incident flux is not conserved. It is  rather amplified or absorbed which can be expressed by the continuity equation as 
\cite{PhysRevA.64.042716,PhysRevA.54.2060}
\begin{equation}
 \frac{\partial j}{\partial x}+\frac{\partial N}{\partial t}=2E_2^{\dagger}(\Gamma+\gamma\sigma_z+\epsilon\sigma_x)E_2.
\end{equation}
Here, $j$ and $N$ denote the energy flux density and the wave intensity density of the EM wave, respectively. The right hand side of the above equation defines the increase or decrease of incident flux density while passing through the region-II. The degree of amplification can be quantified by a coefficient \cite{PhysRevA.64.042716,PhysRevA.54.2060} as
\begin{equation}
 \alpha(\theta)=\frac{j_{t}-j_{r}}{j_{i}} -1=\frac{2}{v\cos\theta}\int_{0}^{L_x}E_2^{\dagger}(\Gamma+\gamma\sigma_z+\epsilon\sigma_x)E_2 dx,
\end{equation}
where $j_{i}$, $j_{r}$ and $j_{t}$ are the incident, reflected, and transmitted flux densities, respectively. 

Using the continuity condition for the wave function across the interfaces at $ x=0$ and $x=L_x$, 
one has
\begin{align}\label{eq1}\nonumber
 &n\chi_{+} + r\chi_{-} = a\psi_{+}+ b\psi_{-},\\
&a\psi_{+}e^{-\kappa L_x}+b\psi_{-}e^{\kappa L_x}= t\chi_{+}e^{ik_xL_x}e^{\epsilon L_x/v}.
\end{align}
The above two equations can be solved to obtain the transmission probability $T=tt^{*}/n^2$ as
\begin{equation}\label{trans}
 T(\theta)=\frac{[v\kappa\cos\theta /\sinh(\kappa L_x)]^2 e^{-2\epsilon L_x/v}}{\left[\Gamma+v\kappa\cos\theta\coth(\kappa L_x)\right]^2+\Omega^2\sin^4\theta}~~~~.
\end{equation}
The reflection probability $R = rr^{*}/n^2$ is given by
\begin{equation}\label{refl}
 R(\theta)=\frac{\Gamma^2\sin^2\theta+(\gamma\cos\theta+\Omega\sin\theta)^2}{[\Gamma+v\kappa\cos\theta\coth(\kappa L_x)]^2+\Omega^2\sin^4\theta}.
\end{equation}  
Taking the difference of energy flux between left and right regions one notes $\alpha(\theta)= T(\theta) + R(\theta) -1$. 

 \begin{figure}[t]
 \centering
\begin{minipage}[t]{.5\textwidth}
  \hspace{-.45cm}{ \includegraphics[width=.5\textwidth,height=4cm]{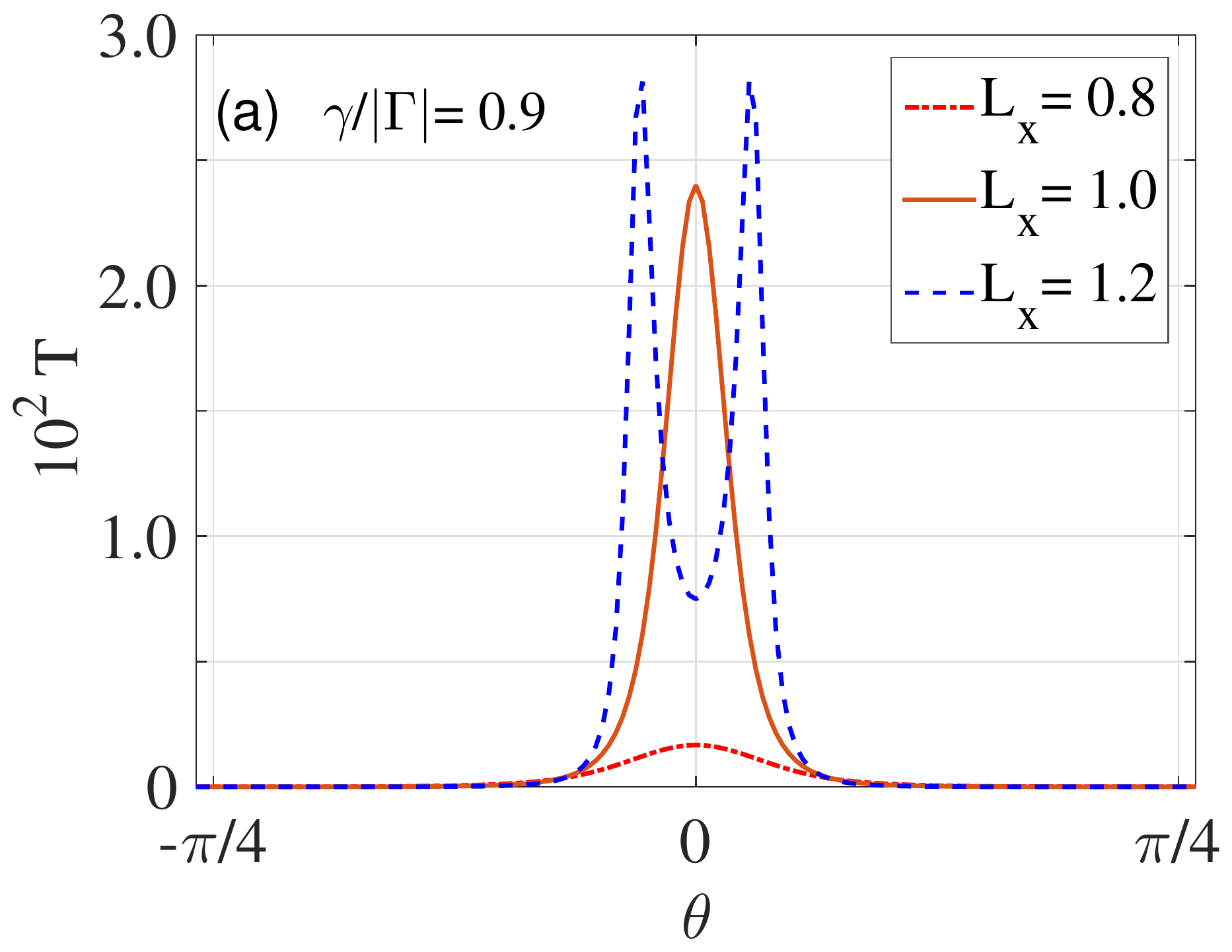}}
  \hspace{-0.1cm}{ \includegraphics[width=.5\textwidth,height=4cm]{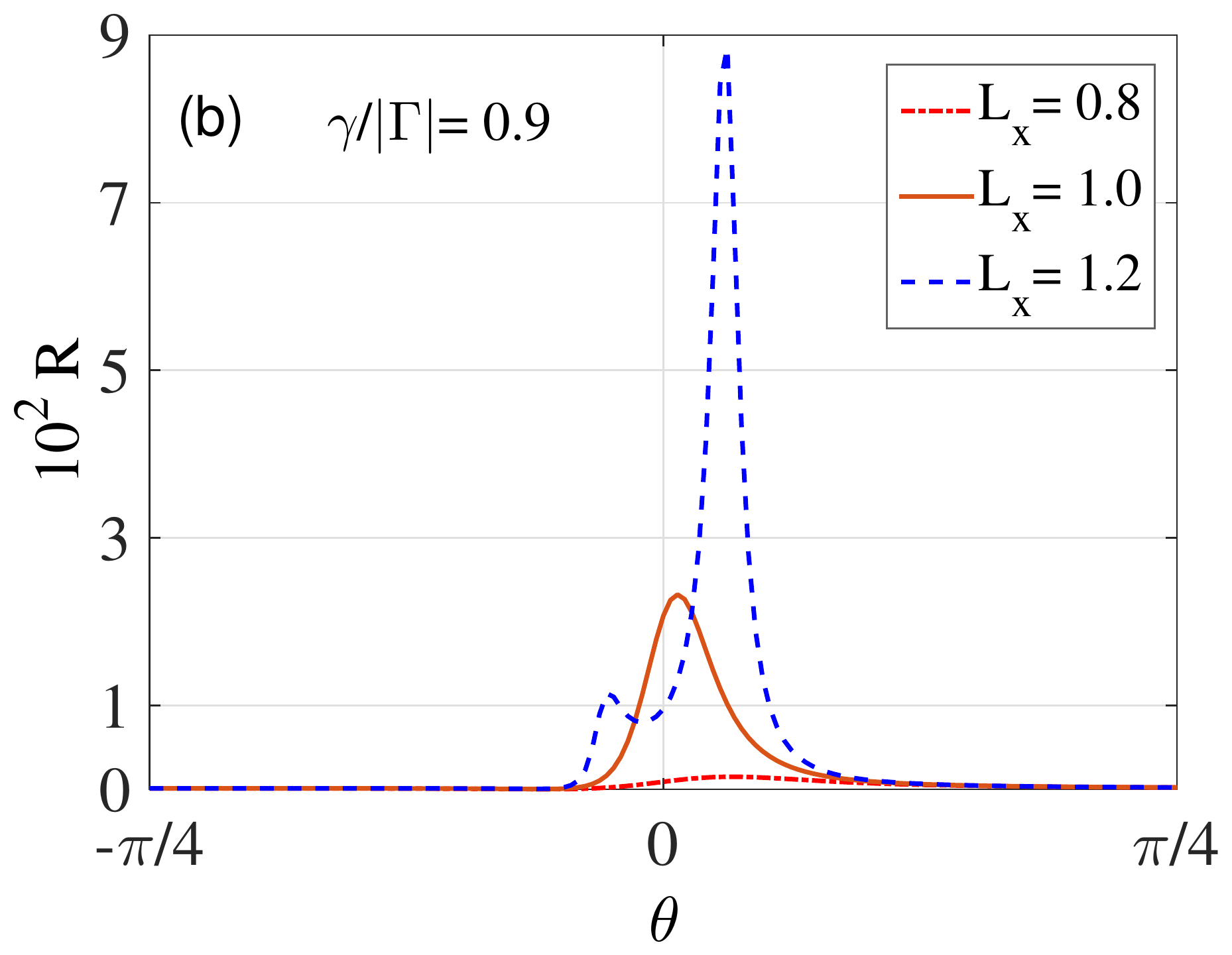}}
\end{minipage}
 \caption{(a) Transmission and (b) reflection probabilities as a function of the angle of incidence $\theta$ under a amplification background. 
 The different lengths are taken in units of $L_{cr}=v/|\Gamma|$ and $\Omega/|\Gamma| = 5$ and $\epsilon=0$. The case with negative $\gamma$ is described by the mirror symmetric plots.}
 \label{TR1}
 \end{figure}
%%%%%%%%%%%%%%%%%%%%%%%%%%%%%%%%%%%%%%%%%%%%%%%%%%%%%%%% conductance for different omega and length%%%%%%%%%%%%%%%%%%%%%%%%%%%%%%%%%%%%%%%%%
Let us discuss the behavior of the transmission and reflection probabilities. 
At $\gamma=0$ and normal incidence of light, the reflection probability is zero, while the transmission through amplifying media increases compared to unity, respectively, as  
$T(0)=e^{2(|\Gamma|-\epsilon) L_x/v}$. The backscattering effect of $\gamma$ can be seen in the appearance of the resonance in transmission probability at the 
normal angle of incidence. This is in contrast to the electron Klein tunneling problem in graphene, \cite{katsnelson2006chiral, Klein_review}. 
Indeed, in the amplifying media at $\Gamma <0$ one obtains
\begin{equation}\label{Threshold}
\tanh\left(\frac{L_x |\Gamma| }{v}\sqrt{1-\gamma^2/\Gamma^2}\right) = \sqrt{1-\gamma^2/\Gamma^2}.
\end{equation}
The solution of the above equation describes the situation of the generator threshold, at which the region-II acts as a source of the radiation in the absence of the incident field, i.e at $n=0$.
At $|\Gamma|\gg|\gamma|$ the threshold length of the system logarithmically diverges with the decrease of $|\gamma|$ as $L_{x,cr} \approx (v/|\Gamma|)\ln|2\Gamma/\gamma|$.
While at $\sqrt{\Gamma^2-\gamma^2}L_x/v \ll 1$ one has $L_{x,cr} \approx v/|\Gamma|$ and the transmission probability at the normal angle of incidence is enhanced as 
\begin{equation}
T(0) = \frac{e^{-2\epsilon/|\Gamma|}}{(1- |\Gamma|L_x/v )^2}. 
\end{equation}
The reflection probability at the normal angle of incidence is strongly increased at the vicinity of threshold 
$
R(0) = (\Gamma/\gamma)^2 /(1- |\Gamma|L_x/v )^2
$,
which is opposite to the reflectionless Klein tunneling through graphene lattice in Hermitian case. The system becomes a generator if $L_x > L_{x,cr}$, so that there are finite solution for $R(0)$ and $T(0)$ at vanishing incident radiation. 

\begin{figure}[t]
 \centering
\begin{minipage}[t]{.5\textwidth}
  \hspace{-.45cm}{ \includegraphics[width=.5\textwidth,height=4cm]{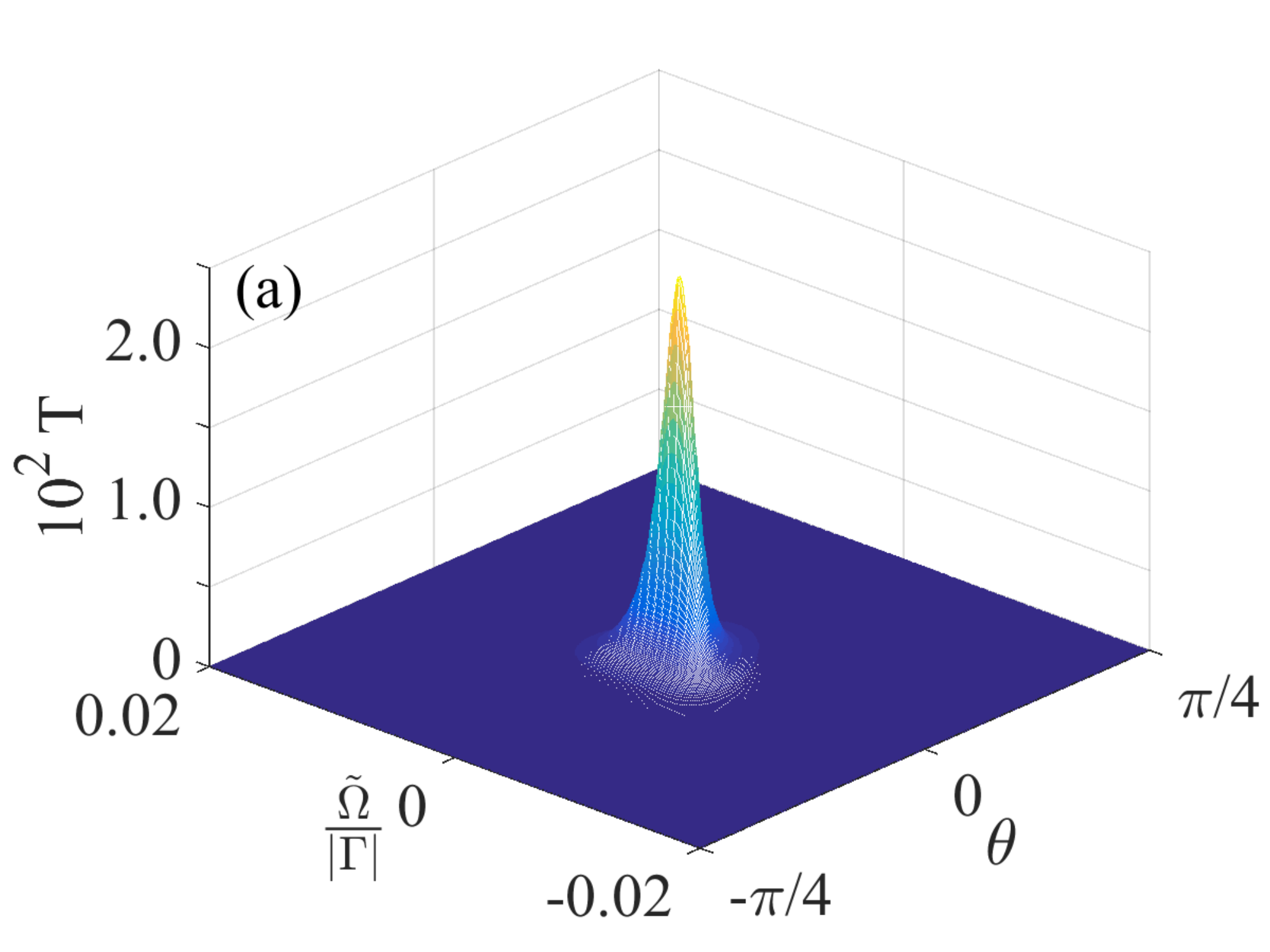}}
  \hspace{-0.1cm}{ \includegraphics[width=.5\textwidth,height=4cm]{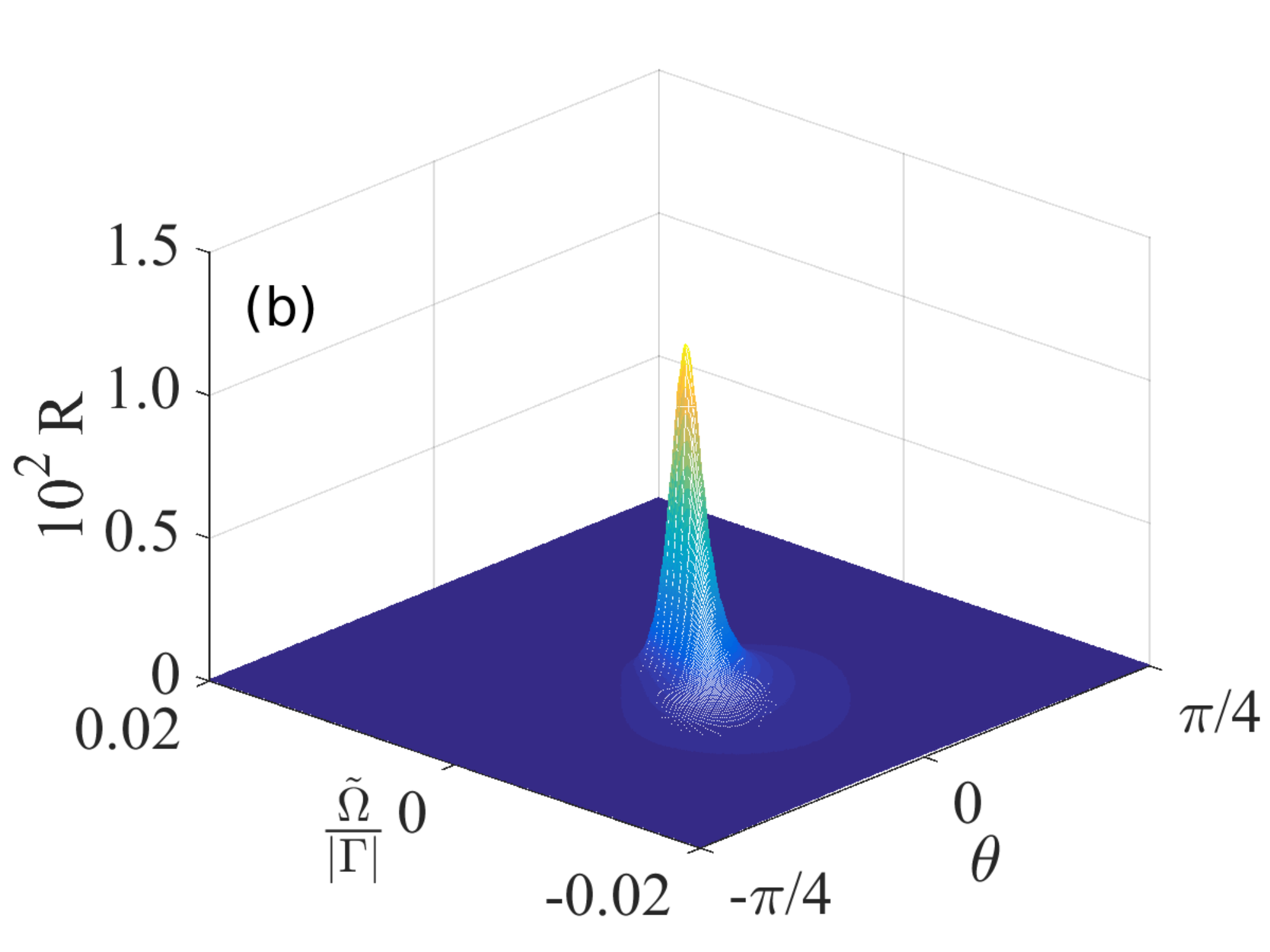}}
\end{minipage}
  \caption{ (a) Transmission and (b) reflection probabilities are shown in the plane of ($\tilde{\Omega}/|\Gamma|-\theta$) at the length $L_{cr}=v/|\Gamma|$. All other parameters are same as in Fig.~(\ref{TR1}). }
 \label{contour}
 \end{figure}

We plot the transmission and reflection probabilities in Fig.~(\ref{TR1}) under an amplification background at the vicinity of the critical length.
Both $T(\theta)$ and  $R(\theta)$ are enhanced at the critical length as show in Fig.~(\ref{TR1}a) and Fig.~(\ref{TR1}b), respectively. 
Moreover, notice that $R(\theta)$ is not symmetric with respect to the angle of incidence due to the term $i\gamma\sigma_z$ in the Eq. \eqref{non_Hermitian_Dirac}, which indicates that the reflected wave picks up additional phase from the small imbalance in loss/gain between two waveguides. This asymmetry in reflection probability is the direct manifestation of inversion symmetry breaking. The asymmetry decreases with reducing the length $L_x$ and the frequency $\Omega$, which can be seen from Eq. \eqref{refl}. Changing the sign of $\gamma$ leads to a mirror symmetric plot of the angle dependence of reflection probability. 
So far we study the case $\tilde{\Omega}=0$. We note that the deviation of $\tilde{\Omega}$ from the zero value inside the amplifying region suppresses the amplitude of the resonance as 
shown in the plots of $T(\tilde{\Omega},\theta)$ and $R(\tilde{\Omega},\theta)$ in Fig.~(\ref{contour}).

%%%%%%%%%%%%%%%%%%%%%%%%%%%

Finally, the amplitude of the transmission probability is sensitive to the direction of incident wave due to complex term $\epsilon$, \cite{Longhi_robust, PhysRevB.92.094204}. 
Note that expressions in Eq. \eqref{trans} is obtained for the wave incident from the left side. 
To find the transmission probability of the wave incident from the right, one has to substitute $e^{-2\epsilon L_x/v} \rightarrow e^{+2\epsilon L_x/v}$ in Eq. \eqref{trans}. 
At $|\epsilon| L_x/v >1$, one of the two scattering processes shall be exponentially suppressed.  
Although the resonance condition Eq. \eqref{Threshold} is independent on $\epsilon$. 

The peak in $T(\theta)$ and $R(\theta)$ gets split into two with the further increase of length beyond the critical value.
This case can not be described by the solution of the linear equation and the saturation has to be taken into account.
The dielectric constant depends on the photon flux density
$
i\Gamma \rightarrow i\Gamma(1- E^\dag \eta E),
$
where the components of the matrix $\eta$ describe saturation of the stimulated emission, \cite{Konotop_RevModPhys, Review_photonics_2018}.
Let one keep $\gamma$ and $\epsilon$ fixed and tune the pump $|\Gamma|$ over the threshold value $|\Gamma_{cr}|$ defined by Eq. \eqref{Threshold}. 
Above the threshold, the field in the middle region is given by $E_2 = a \psi$, with
\begin{equation}
\psi = \left\{ \psi_{+}e^{-\kappa x} + \frac{|\Gamma|}{\gamma}\left[\sqrt{1-\gamma^2/\Gamma^2} -1\right] \psi_{-}e^{\kappa x}\right\} e^{ik_y y-\epsilon x/v}, 
\end{equation}
where the weak nonlinearity $0<E^\dag \eta E < |\gamma/\Gamma|$, determines the amplitude of the solution.
\begin{equation}
|a|^2 = (1- |\Gamma_{cr}/\Gamma|) \frac{ \int_0^{L_x}dx |\psi|^2}{\int_0^{L_x}dx|\psi|^2 (\psi^\dag \eta \psi)},
\end{equation}
where $|\Gamma/\Gamma_{cr}|$ is the coefficients of the excess of the pump over the threshold value.

\section{Thin scattering region}
\begin{figure}[t]
 \centering
\begin{minipage}[t]{.5\textwidth}
  \hspace{-.3cm}{ \includegraphics[width=.5\textwidth,height=4cm]{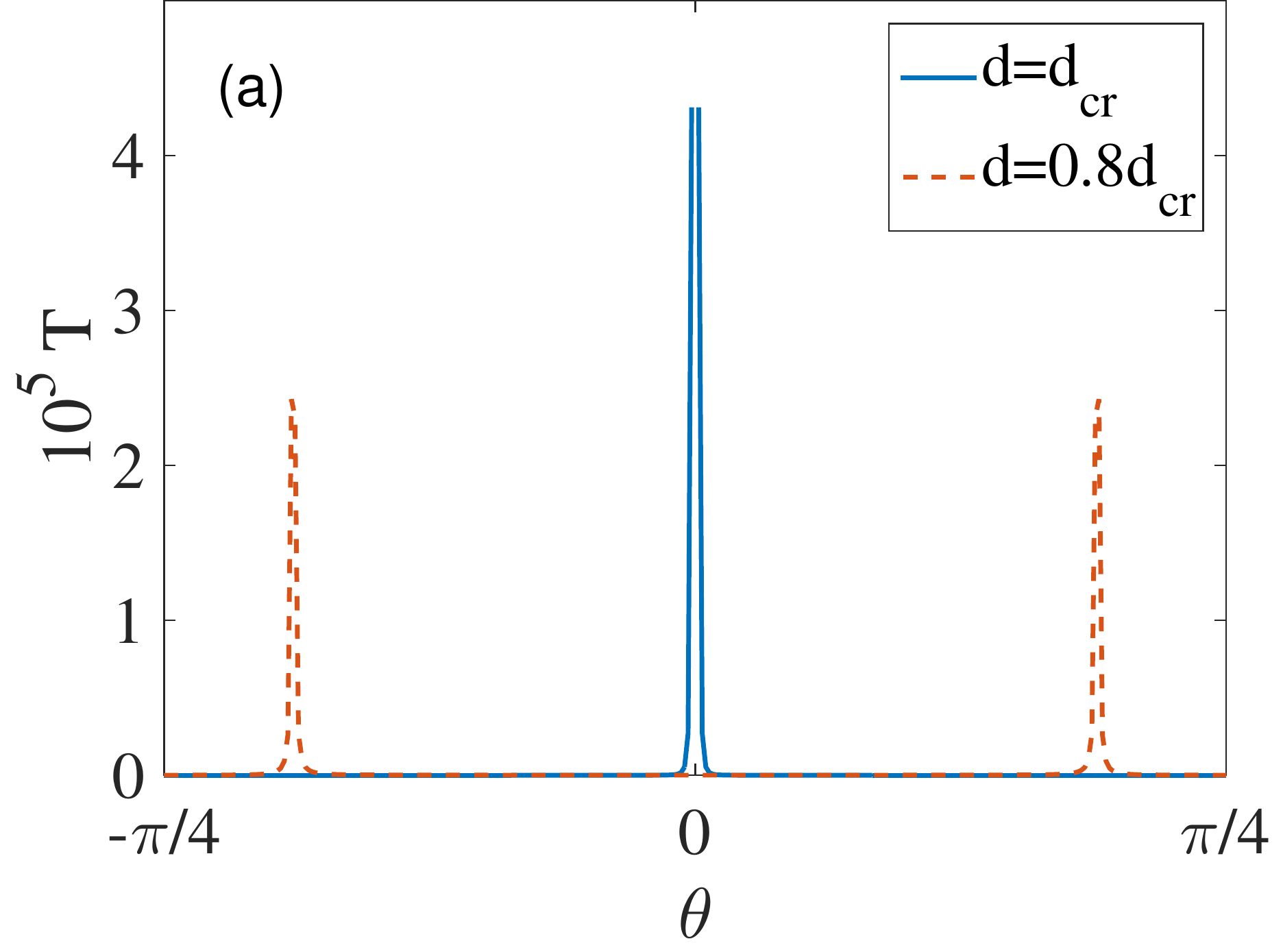}}
  \hspace{-0.1cm}{ \includegraphics[width=.5\textwidth,height=4cm]{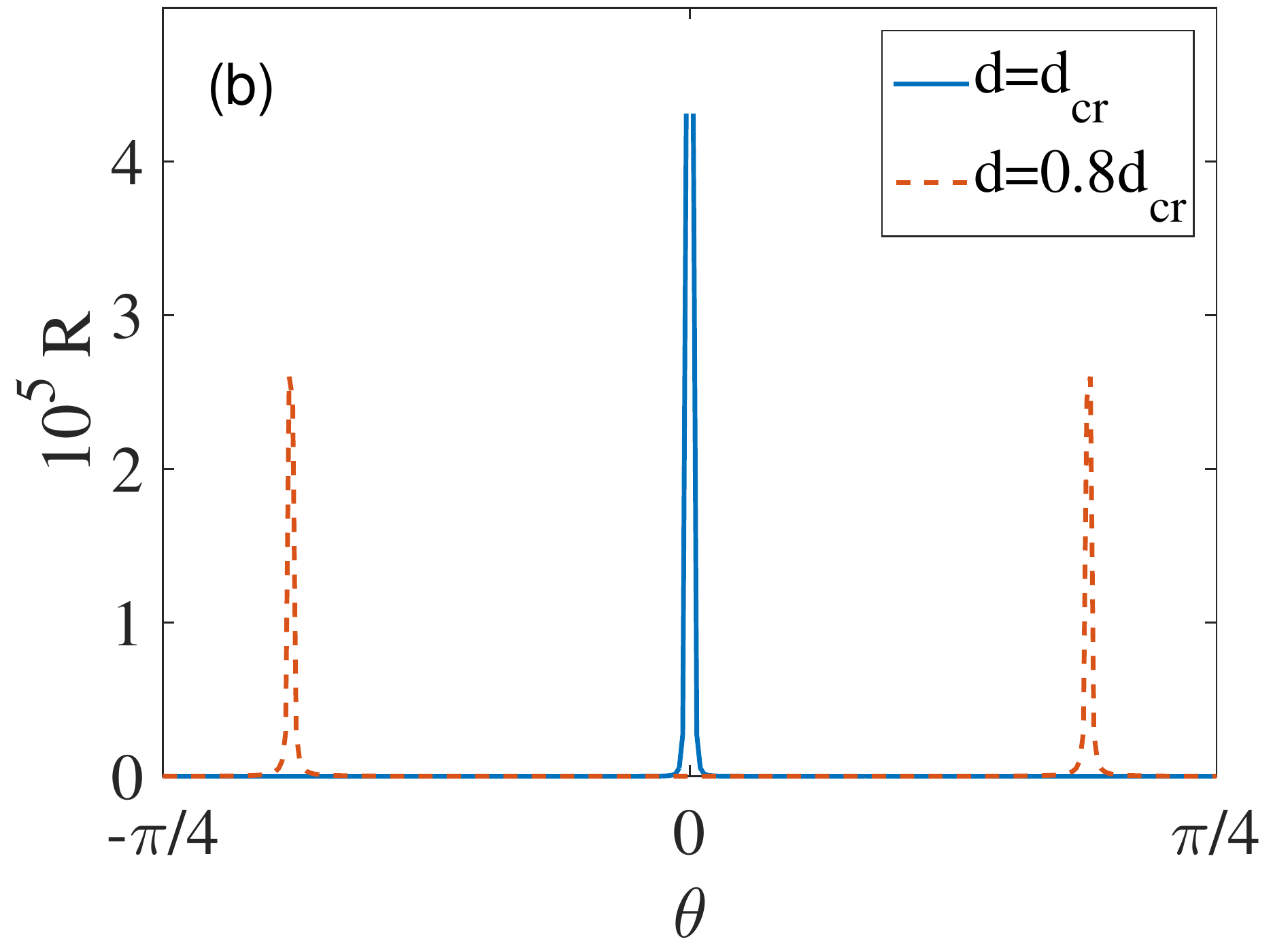}}
\end{minipage}
  \caption{ Resonance in (a) transmission and (b) reflection probabilities. Here, $d_{cr}\sim v/\mid \Gamma\mid$ and all other parameters are same as in Fig.~(\ref{TR1}).}
 \label{thin}
 \end{figure}

Let us also comment on the case when the amplifying region is very thin $L_x\rightarrow d$ and strong so that $|\Gamma|,|\gamma|\gg |\tilde{\Omega}|$, where $d=v/\Lambda$ is the short rang cut-off with the frequency width $\Lambda$, within which the model of Dirac-like spectrum is valid. 
This can be realized by plugging all the waveguides along the $y$-direction at $x=0$. Here the transmission probability is given by
\begin{equation}\label{trans1}
 T(\theta)= \left[\frac{v\kappa\cos\theta/\sinh(d\kappa)}{\Gamma+v\kappa\cos\theta\coth(d\kappa)} \right]^2e^{-2\epsilon d/v},
\end{equation}
and the reflection probability (for $\epsilon=0$) can be obtained as
\begin{equation}
 R(\theta)=\frac{\Gamma^2\sin^2\theta+\gamma^2\cos^2\theta}{[\Gamma+v\kappa\cos\theta\coth(d\kappa)]^2}.
\end{equation}
The resonance survives in the thin limit $L_x\rightarrow d$, provided the condition $d|\Gamma|/v\sim 1$ is satisfied. 
However, apart from the normal incidence, the resonance can also be seen for any arbitrary angle, provided $v\kappa \cos\theta_c=\Gamma \tanh(d\kappa)$ for the amplifying background.
The amplitude of the solution inside the thin region can also be obtained by incorporating the above mentioned limits.
We plot the  transmission and reflection probabilities in Fig.~(\ref{thin}), which show that the  resonance can occur at normal angle of incidence as well as away from the normal incidence for different parameters. It is also interesting to note that the width and height of the resonance peak are much sharper in contrast to the wide junction case [see Fig.~(2)], which is attributed to the extra term $\Omega^2\sin^4\theta$ in the denominator of eq.~(9). It can also be clearly seen that the effects of the imaginary gauge field ($\epsilon$) remains
insensitive to the thickness of the junction.
%%%%%%%%%%%%%%%%%%%%%%%%%%%
The reflection probability is fully suppressed at normal incidence without inversion symmetry breaking, $\gamma=0$. 

\section{Discussion}
We shall also comment on the propagation of the radiation through the photonic honeycomb lattice with dissipation at $\Gamma > |\gamma|, |\epsilon|$.
Suppose that the incident light is coming on the dissipative media from both regions I and III.
The resonance condition at which the amplitude of the reflecting waves vanishes
can be understood from the fact that the divergence of $T(0)$ in Eq.~(\eqref{trans}) depends on the sign of $\Gamma \cos\theta$. Note that the spinor of the wave incident on the media from the right side, acquires a sign reversal of $\cos\theta\rightarrow\cos(\pi-\theta)$. Hence the pole in the amplitude of the incoming waves takes place at positive $\Gamma$. In this case the system acts as an absorber of the radiation above the threshold Eq. \eqref{Threshold} with absorption coefficient $-1<\alpha<0$.

The interplay of non-Hermiticity and topology in $2$D PhL have been extensively considered in the last decade \cite{Review_photonics_2018, torres} including the $\mathcal{PT}$-symmetric case with balanced loss and gain, see for example \cite{PhysRevB.98.165129, kremer2019demonstration}. However, resonances in the ballistic transmission of light through a honeycomb PhL with such background have been overlooked so far.
We consider a honeycomb PhL under a uniform amplification or dissipation to each waveguides, {\it i.e.}, a broken $\mathcal{PT}$-symmetric lattice. The PhL can provide emission or perfect absorption of light above a certain threshold length. Here the resonant feedback is due to inversion symmetry breaking term in the Dirac-like wave-equation. We should also mention that the imaginary gauge field has no impact the generator threshold in our model. Although it leads to an exponential enhancement or suppression of the transmission probability, which is sensitive to the direction of incident light. The effect of imaginary gauge field remains robust to the thickness of the amplifying junction. 

It is also instructive to consider the effect of multiple light scatterings at a point-like defect, which is described by a potential  
$i\gamma \sigma_z\delta(\mathbf{r}-\mathbf{r}_i)$ at a position $\mathbf{r}=\mathbf{r}_i$. The Green function of Eq. \ref{non_Hermitian_Dirac} at $\epsilon=0$ and under the substitution 
$i\gamma \sigma_z \rightarrow i\gamma \sigma_z\delta(\mathbf{r}-\mathbf{r}_i)$ in presence of the single
impurity satisfies an equation
\begin{equation}\label{dyson}
[\tilde{\Omega} + i\Gamma +i\gamma \sigma_{z} \delta(\mathbf{r}-\mathbf{r}_i)+iv \boldsymbol{\sigma}\boldsymbol{\partial}_r ]\mathcal{G}(\bf {r,r'})=\delta({\bf r-r'}),
\end{equation}
where the Green function is normalized to $\omega_D^2/c$.  To obtain the possible localized or trapped states, we solve the impurity scattering problem by computing the poles of the $T$-matrix, which can be found from the equation
$ {\rm det}[1+i\gamma \sigma_{z}\mathcal{G}(\mathbf{r}_i,\mathbf{r}_i)]=0 $, where the bare Green function is given by 
\begin{eqnarray}\label{green}
 \mathcal{G}(\mathbf{r}_i,\mathbf{r}_i)=-\frac{\tilde{\Omega} + i\Gamma}{4\pi v^2}\ln\frac{\Lambda^2}{(\Gamma-i\tilde{\Omega})^2}.
\end{eqnarray}
%with the frequency cut-off $\omega_0$, above which the model of Dirac-like spectrum is not valid.
%The Green function of Eq. system can be written as
%\begin{equation}
% \mathcal{G}(\omega;{p})=\frac{\tilde{\omega} + i\Gamma-i\gamma \sigma_z+v \boldsymbol{\sigma}\cdot\mathbf{p}}{(\tilde{\omega} + i\Gamma)^2+\gamma^2 - v^2p^2},
%\end{equation}
One obtains two equations for the poles 
\begin{equation}
\left(\Gamma-i\tilde{\Omega} \right)\ln\frac{\Lambda^2}{(\Gamma-i\tilde{\Omega})^2} =\pm \frac{4\pi v^2}{\gamma}.
\end{equation}
At the Dirac point, in the limit of large potential $|\gamma| \gg v^2/\Lambda$, the poles for both amplifying and dissipative media are given by $\Gamma =\pm 2\pi v^2/\gamma\ln|\gamma\Lambda/2\pi v^2|$.

Finally, besides the resonant feedback, it has been long known that disorder might provide a feedback for generation of light, \cite{letokhov1968generation}. 
The properties of the random laser has been extensively studied, for a review see \cite{wiersma2008physics}.
It would be interesting to extend the above results to the light propagation through disordered amplifying photonic honeycomb lattice.

\section{Conclusion}
To conclude, we investigate the propagation of the electromagnetic wave through an amplifying region in a photonic honeycomb lattice. We reveal that there exists a critical length of the amplifying region for which the transmission and reflection probabilities of the wave diverge at normal angle of incidence. The condition for the generator threshold is determined by the parameters associated to the lattice structure and amplification  background. 
The amplification is sensitive to the direction of incident wave in presence of an imaginary gauge field in the amplifying region. We also comment on the resonant states at a thin scattering region. The possible existence of localized states in presence of a single impurity is also discussed. Our investigation on resonance might be realized in realistic set-ups, designed in honeycomb photonic or plasmonic lattices \cite{kremer2019demonstration,PhysRevLett.122.013901}, where such feedback has been artificially imparted.

{\it Acknowledgements:}
This work is supported by the Academy of Finland Grant No. 308339. 
A.A.Z. is grateful to the hospitality of the Pirinem School of Theoretical Physics. 

\bibliography{bibfile_PhC}
\end{document}